\documentclass[namedreferences]{kluwer}
\usepackage[applemac]{inputenc}
\usepackage{graphicx}

\newcommand{\ave}[1]{\left \langle {#1} \right \rangle}
\newcommand{\abs}[1]{\left \vert {#1} \right \vert}

\begin{document}

\begin{opening}
\title{Rate equation theory of sub-Poissonian laser light}

\author{J. \surname{Arnaud}\email{arnaudj2@wanadoo.fr}}
\institute{Mas Liron, F30440 Saint Martial, FRANCE}

\runningtitle{Rate equation theory of sub-Poissonian laser light}
\runningauthor{Arnaud J.}

\begin{abstract}
Lasers essentially consist of single-mode optical cavities containing
two-level atoms with a supply of energy called the pump and a sink of
energy, perhaps an optical detector.  The latter converts the light
energy into a sequence of electrical pulses corresponding to
photo-detection events.  It was predicted in 1984 on the basis of
Quantum Optics and verified experimentally shortly thereafter that
when the pump is non-fluctuating the emitted light does not fluctuate
much.  Precisely, this means that the variance of the number of
photo-detection events observed over a sufficiently long period of
time is much smaller than the average number of events.  Light having
that property is said to be ``sub-Poissonian''.  The theory presented
rests on the concept introduced by Einstein around 1905, asserting
that matter may exchange energy with a wave at angular frequency
$\omega$ only by multiples of $\hbar\omega$.  The optical field energy
may only vary by integral multiples of $\hbar\omega$ as a result of
matter quantization and conservation of energy.  A number of important
results relating to isolated optical cavities containing two-level
atoms are first established on the basis of the laws of Statistical
Mechanics.  Next, the laser system with a pump and an absorber of
radiation is treated.  The expression of the photo-current spectral
density found in that manner coincides with the Quantum Optics result. 
The concepts employed in this paper are intuitive and the algebra is
elementary.  The paper supplements a previous tutorial paper
\cite{arnaudOQE} in establishing a connection between the theory of
laser noise and Statistical Mechanics.
\end{abstract}


\classification{PACS numbers}{42.55.Ah,  42.50.Ar, 42.55.Px, 42.50.Lc}

\end{opening}

\section{Introduction}

The purpose of this paper is to present a derivation of the essential
formulas relating to sub-Poissonian light generation in a simple and
self-contained manner.  This is done on the basis of a theory in which
the light field enters through its energy.  Only single-mode cavities
incorporating emitting and absorbing atoms are considered.  Some
Quantum-Optics effects \cite{elk} that become inconspicuous when the
number of atoms is large are neglected.  The paper does not assume
specialized knowledge from the reader, but some understanding of
general concepts relating to random variables (mean, variance) and
stationary random functions of time (spectral densities)
\cite{papoulis}, may be useful.
 
\begin{figure}
    \centering
    \includegraphics[scale=0.7]{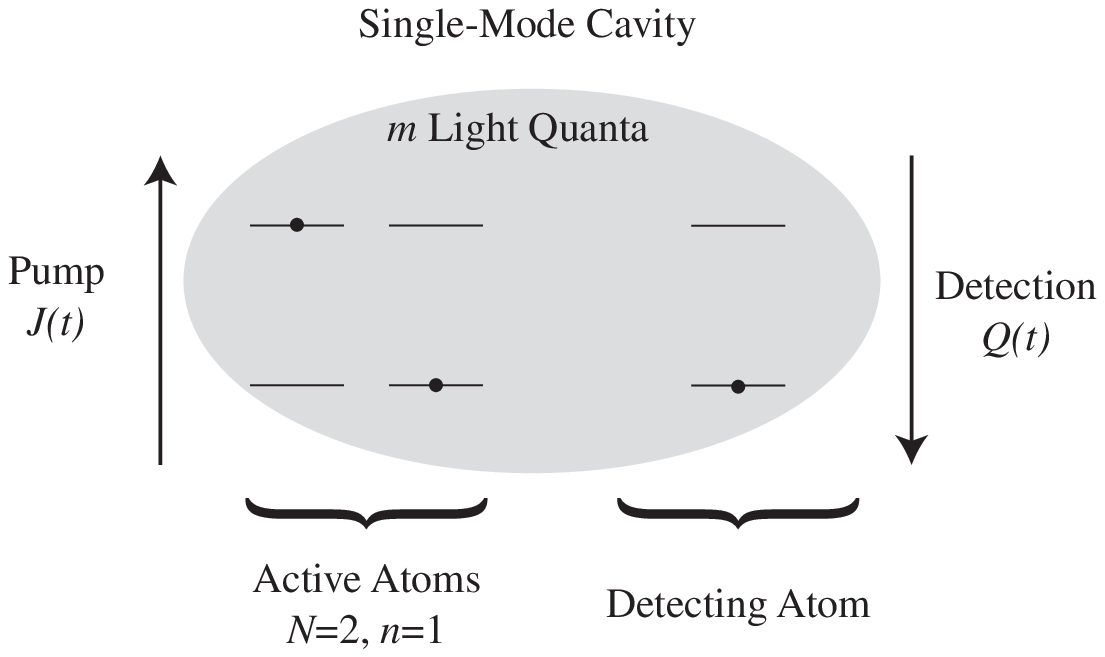}
	\caption{Schematic representation of a laser.  The single-mode
	cavity (shown as a gray oval) incorporates active two-level
	atoms (only two are shown on the left, one in the upper state
	and one in the lower state) and absorbing (detecting) atoms
	(only one is shown on the right).  The pump raises active
	atoms from their lower to upper states at some prescribed rate
	$\mathcal{J}$.  Photo-detection events occur at rate
	$\mathcal{Q}$.  At low frequencies,
	${\mathcal{Q}}(t)={\mathcal{J}}(t)$.}
    \label{fig:1}
\end{figure}

Laser light possesses high degrees of directivity and
monochromaticity.  The intensity fluctuations, though relatively
small, are of practical significance in some applications:
transmission of information by means of optical pulses, measurement of
small attenuations, or interferometric detection of gravitational
waves.  Lasers essentially consist of single-mode optical cavities
containing resonant atoms with a supply of energy called the pump and
a sink of energy, presently viewed as an optical detector (see Figure
\ref{fig:1}).  The latter converts light into a series of identical
electrical pulses, referred to as ``photo-detection events''
\cite{koczyk}.  If the events are independent of one another, the
light impinging on the detector is said to be ``Poissonian'', and the
photo-current fluctuations are at the ``shot-noise level''.  But under
some circumstances detection events occur more regularly, in which
case the light is said to be ``sub-Poissonian''.  Light of that nature
has been first observed by \inlinecite{short}.  Subsequently,
\inlinecite{yamamoto} performed a remarkable series of experiments on
laser diodes.  They observed a reduction by up to a factor of 10 below
the shot-noise level.  The correlation between the number of
upper-state atoms in the cavity and photo-detection events has also
been measured \cite{richardson}.
 
A single-mode optical cavity resonating at angular frequency $\omega$
may be modeled as an inductance-capacitance ($L,C$) circuit with
$LC\omega^2=1$.  The active atoms, located between the capacitor
plates, interact with a spatially uniform optical field through their
electric dipole moment (see, for example, \inlinecite{milonni}).  In
this idealized model, the field angular frequency ``seen'' by the
atoms is equal to $\omega$~\endnote{In the present idealized model, no
momentum is transfered between the field and the atoms, so that the
Maxwellian atomic-velocity distribution is undisturbed.  An advanced
relativistic treatment can be found in \inlinecite{benyaacov}.}.  The
2-level atoms (with the lower level labeled ``1'' and the upper level
labeled ``2'') are supposed in this paper to be resonant with the
field.  This means that the atomic levels 1 and 2 are separated in
energy by $\hbar\omega$, where $\hbar$ denotes the Planck constant
divided by $2\pi$.  The energy unit is taken equal to $\hbar\omega$,
for simplicity.  The pumping rate $\mathcal{J}$, defined as the number
of atoms raised to their upper state per unit time, is supposed to be
a prescribed function of time, i.e., to be independent of the laser
dynamics.  This condition is actually achieved in the case of
semiconductors with the help of high-impedance electrical sources.  It
has been shown further \cite{khazanov,ritsch,ralph} that
sub-Poissonian output light statistics should be obtainable also from
four-level atom lasers with incoherent optical pumping~\endnote{In the
case of 4-level atom lasers (with levels labeled from 0 to 3, the
working levels being those labeled 1 and 2), strong optical pumping
resonant with levels 0 and 3 provides a constant probability that
electrons in level 0 be transfered to level 3 per unit time, and
(almost) the same probability that electrons in level 3 be transfered
to level 0.  The detected-light fluctuations may be sub-Poissonian at
zero baseband frequency.  Precisely, the spectral density is half the
shot-noise level under ideal conditions (negligible spontaneous decay,
negligible optical loss, and quick decay from level 3 to level 2). 
This desirable behavior results from level 0 population
fluctuations.}.  This important observation escaped the attention of
previous laser theorists.  For such four-level lasers the expressions
obtained from the present theory are identical to those reported in
these references \cite{chusseau}.  The output light fluctuates at only
one third of the shot-noise level.

The light-energy absorber is modeled as an ideal optical detector that
generates a series of identical electrical pulses, the energy lost by
the field being entirely dissipated in the detector load.  Unlike
active atoms, detector atoms remain most of the time in the ground
state, with quick non-radiative decay after an excitation event.  For
the sake of conceptual clarity, the absorber of radiation is supposed
to be located within the optical cavity as in the \inlinecite{sargent}
classical text-book~\endnote{In our model, detection (linear
absorption) is supposed for the sake of simplicity to occur within the
optical cavity.  But no difference of behavior is observed when a
laser beam is absorbed externally without reflection, rather than
internally, at the same average rate.  It is therefore expected that
the present theory be applicable to reflectionless external detectors. 
In open-space configurations, legitimate questions could be raised in
connection with the law of causality.  It is therefore important to
emphasize that only single-mode cavities are considered in the present
paper, and that the concept of propagating light is not relevant. 
Note further that the theory would hold just as well if the optical
resonator were replaced, for example, by an acoustical resonator at
the same frequency.  The wave is localized, and its physical nature is
unimportant.}.  The main purpose of this paper is to derive an
expression of the photo-current spectral density, particularly in the
limit of small Fourier (or ``baseband'') frequencies \endnote{In
Optical Communications, $\Omega$ is often called ``baseband'' angular
frequency and $\omega$ ``carrier'' angular frequency.  It is a common
practice to consider only \emph{positive} baseband frequencies, so
that factors of 2 may arise as one goes from the Physics to the
Engineering literature.  The time dependence at baseband frequencies
is denoted in this paper by: exp$(j\Omega t)$.  To define the
photocurrent spectral density, consider a particular (experimental or
computer-generated) run lasting from $t=0$ to $\tau _{m}$, the
photodetection events occurring at times $t_{1},t_{2},\ldots, t_{i},
\ldots $.  The detection rate ${\mathcal{Q}}(t)$ is defined as the sum
over $i$ of $\delta (t-t_{i})$, where $\delta(.)$ denotes the Dirac
distribution.  The detector electrical current, if desired, would be
obtained by multiplying ${\mathcal{Q}}(t)$ by $e$, the absolute value
of the electron electrical charge.  The (real, nonnegative) spectral
density $S_{\Delta Q}(\Omega)$ of $\Delta Q\equiv Q(t)-\ave{Q}$ is
defined as: $\ave{\abs{\sum \exp ( j \Omega t_{i})}^2} / \tau_{m}$ ,
where brackets denote an average over many runs, and $\Omega =2\pi
n/\tau_{m}$, with $n=1,2\ldots$.  This expression is accurate if the
measurement time $\tau_{m}$ is sufficiently large.  In the special
case where the photo-detection events are independent of one another,
and uniformly distributed (uniform Poisson process), we have
$S_{\Delta Q}(\Omega)=\ave{Q}$, a relation usually referred to as the
``shot-noise formula''.  The variance of the number of events
occurring during some time T is in that case equal to the average
number of events, see Eq.~(5.37) of \cite{papoulis}.}.

Consider now some theories of sub-Poissonian light generation.  The
main concept of Quantum Optics, introduced by Dirac in 1927, is that
the field in a cavity should be treated as a quantized harmonic
oscillator.  Following Dirac's lead, laser theorists, initially, were
mostly concerned with the state of the cavity field.  But
\citeauthor{golubevJETP}, in \citeyear{golubevJETP}, carefully
distinguished the fluctuations of the field in the cavity from those
of the detection rate.  They further pointed out that when lasers are
driven by non-fluctuating pumps, the emitted light should be
sub-Poissonian.  Similar results were subsequently obtained by
\inlinecite{yamamoto}, see also
\cite{meystre,walls,davidovitch,mandel}.  These authors employed
various approximations of the laws of Quantum Optics, and various
models to describe non-fluctuating pumps.  No approximation is made in
the recent numerical work of \inlinecite{elk}.  But because the
computing time grows exponentially with the number of atoms present in
the cavity, the author is able to give results only for up to 5 atoms. 
Even with few atoms, some typical features of Quantum Optics (the
so-called trapped states) get washed out, so that simplified theories
may be adequate.  \inlinecite{loudon} and \inlinecite{jakeman} have
treated the evolution in time of the number of photons in a laser
amplifier on the basis of a theory in which the optical field is not
explicitly quantized, as is done here.  Such theories are able to
describe sub-Poissonian light when the detecting atoms are included in
the system description~\endnotemark[3].
Accurate noise sources have been obtained by Gordon in 1967 (see
Section 5 of \inlinecite{gordon} entitled ``The generalized Wigner
density and the approximate classical model'') through symmetrical
ordering of operators.  Gordon did not address, however, the case of
non-fluctuating pumps.  Application of the Gordon formalism to
non-fluctuating-pump lasers was made independently in 1987-88 by
\inlinecite{katanaev} and this author \cite{arnaudOL}.  A discussion
is given by \inlinecite{nilsson}.

Theories in which the atoms are quantized but the optical field is not
directly quantized are usually labeled ``semiclassical''.  However,
because this adjective may cause confusion with alternative
theories~\endnote{Many authors attempted to avoid the intricacies of
field-quantization.  But while there is essentially only one quantum
theory, there exist many distinct theories in which the field is
treated in a classical manner.  When the operators entering in the
exact Quantum Optics theory are normally ordered, linearized, and the
operator character of the field is ignored, a theory emerges called in
the Optical-Engineering literature the ``phasor'' theory.  According
to that theory, quantum noise would be caused by the field
spontaneously emitted by atoms in the excited state.  But a detailed
comparison (see Appendix B of \inlinecite{arnaudIEEJ}) shows that the
phasor theory, though plausible on some respects, is unable to explain
the origin of sub-Poissonian light.  On the other hand,
\inlinecite{funk}, \inlinecite{raymer}, and \inlinecite{savage}, note
that ``The observed sub-Poissonian statistics are unexplainable using
classical and semi-classical theories''.  This statement applies to
the usual semiclassical theories in which the absorber is forbidden to
react on the field.  But a key point of the present theory is that
absorbers do react on the field.  The observation that
``Sub-Poissonian statistics are possible only for a non-classical
field'' \cite{mandel} is meaningful in the context of Quantum Optics,
but not in the context of the present theory since the ``state'' of
the field (in the Quantum Optics sense) is not considered.} the
present theory has been called simply: ``classical'', to emphasize
that the light field enters only through its energy, a classical
quantity.  But the reader is warned that the theory is not strictly
classical.  Randomness enters because the atomic transitions obey a
probability law rather than a deterministic law.  Note also that the
expression \emph{rate equation} employed in the title sometimes refer
to the time evolution of \emph{average} quantities, and random
fluctuations are ignored.  In this paper, the expression \emph{rate
equation} is understood in a broader sense as, e.g., in the paper by
\inlinecite{ralph}.

The theory presented rests first of all on the concept introduced by
Einstein early in the previous century asserting that matter may
exchange energy with a wave at angular frequency $\omega$ only by
multiples of $\hbar\omega$ \cite{kuhn}.  The law of energy
conservation in isolated systems entails that, if the matter energy is
quantized, the field energy may only vary by integral multiples of
$\hbar \omega$.  Thus, no independent degree of freedom is ascribed to
the field.  This is in sharp contrast with the Quantum Optics
view-point.  The physical picture that emerges from the present theory
is that laser-light fluctuations are caused by the random
\emph{stimulated} transitions responsible for light emission
\emph{and} absorption~\endnote{During a jump from one state to
another, an atom is in a state of superposition.  But such states need
not be considered explicitly as only global conservation laws are
being employed (quantum jumps are discussed, e.g., in
\inlinecite{greenstein}).  Likewise, the interaction energy that may
exist during a jump needs not be considered explicitly.}.  If the
number of atoms in the upper state is denoted by $n$ and the number of
light quanta~\endnote{The word ``photon'' suggesting that light
consists of tiny particles moving in space should better be avoided in
the context of this paper, see the interesting paper by
\inlinecite{lamb}.} in the cavity by $m$, $n+m$ is a conserved
quantity in isolated systems.  It may vary only in response to the
pump generation rate or the detector absorption rate.

The paper is organized as follows.  It is first observed in Section
\ref{sec:2} that useful information on laser light may be obtained by
considering isolated optical cavities containing atoms in a state of
equilibrium, and using the methods of Statistical Mechanics (for an
introduction to that field, see for instance \cite{schroeder}).

When the total system (matter$+$field) energy is sufficiently large,
the equilibrium state is highly non-thermal.  In fact, the gain
saturation mechanism that characterizes laser operation is as work in
the isolated system as well: whenever the light intensity exceeds its
average value there is a decrease of the number of atoms in the
excited state (through energy conservation), and therefore a reduction
of the total probability that a stimulated emission event will occur
within the next elementary time interval.  This effect prevents the
light intensity from varying much.  The great interest of the laws of
Statistical Mechanics is that they provide important informations
about the equilibrium state without having to consider in detail how
the system evolves in the course of time.  Precisely, we find that the
variance of the number of light quanta in the cavity is half the
average number of quanta, that is, the field statistics is
sub-Poissonian.  This result is of direct practical significance if
the light energy contained in the cavity is allowed to radiate into
free-space at some instant.  The emitted light pulse is indeed
sub-Poissonian, though not fully ``quiet''.

In Section \ref{sec:3}, the equilibrium situation enables us to
recover the Einstein prescription asserting that the probability that
electrons be promoted to upper levels is $Cm$ and the probability that
they be demoted to lower levels is $C(m+1)$, where $m$ denotes the
number of light quanta in the cavity and $C$ a constant proportional
to the Einstein $B$-coefficient of stimulated emission and absorption. 
The equilibrium situation provides the rate at which light quanta
would be absorbed at high Fourier frequencies.

But, in order to obtain accurately the rate at any Fourier frequency,
it is necessary to include explicitly pump fluctuations and the
reaction of the absorbed rate on the field (see Section \ref{sec:4}). 
An appendix clarifies the fact that, even though no entropy is
ascribed to the field in the present theory, the isolated system
entropy increases when some piece of matter is introduced into an
initially empty cavity, as the second law of Thermodynamics requires. 
It is shown that the entropy that Quantum Optics ascribes to
single-mode fields is the difference between the system entropy and
the average matter entropy.

\section{Isolated cavities in a state of equilibrium}
\label{sec:2}

Consider $N$ identical two-level atoms.  For each atom, the zero of
energy is taken at the lower level and the unit of energy at the upper
level (typically, 1 eV).  The atoms are supposed to be coupled to one
another so that they reach a state of equilibrium before other
parameters have changed significantly.  The strength of the atom-atom
coupling, however, needs not be specified further.  The atoms are
supposed to be at any time in either the upper or lower state.  The
number of atoms that are in the upper state is denoted by $n$, and the
number of atoms in the lower level is therefore $N-n$.  According to
our conventions, the atomic energy is equal to $n$.  Its maximum value
$N$ occurs when all the atoms are in the upper state.  There is
population inversion when the atomic energy $n>N/2$.

The statistical weight $W(n)$ of the atomic collection is the number
of distinguishable configurations corresponding to some total energy
$n$.  For two atoms ($N=2$), for example, $W(0)=W(2)=1$ because there
is only one possible configuration when both atoms are in the lower
state $(n=0)$, or when both are in the upper state $(n=2)$.  But
$W(1)=2$ because the energy $n=1$ obtains with \emph{either one} of
the two (distinguishable) atoms in the upper state.  For $N$ identical
atoms, the statistical weight (number of ways of picking up $n$ atoms
out of $N$) is (see \inlinecite{papoulis}, p.  58)
\begin{equation}
    W(n)=\frac{N!}{n!(N-n)!} .
    \label{Wofn}
\end{equation}
Note that $W(0)=W(N)=1$ and that $W(n)$ reaches its maximum value at
$n=N/2$ (supposing $N$ even), with $W(N/2)$ approximately
equal to $2^{N}\sqrt{2/\pi N}$. Note further that
\begin{equation}\label{Z}
    Z\equiv \sum_{n=0}^{N}W(n)=2^{N} .
\end{equation}

Consider next an isolated single-mode optical cavity (see Figure
\ref{fig:1} without the pump and the detector), containing $N$
resonant two-level atoms.  The atoms perform jumps from one state to
another in response to the optical field so that the number of atoms
in the upper state is a function $n(t)$ of time.  If $m(t)$ denote the
number of light quanta at time $t$, the sum $n(t)+m(t)$ is a conserved
quantity (essentially the total atom+field energy).  Thus, $m$ jumps
to $m-1$ when an atom in the lower state gets promoted to the upper
state, and to $m+1$ in the opposite situation.  If $N$ atoms in their
upper state are introduced at $t=0$ into the empty cavity ($m=0$),
part of the atomic energy gets converted into field energy as a result
of the atom-field coupling and eventually an equilibrium situation is
reached.  The basic principle of Statistical Mechanics asserts that
all states of isolated systems are equally likely.  Accordingly, the
probability $P(m)$ that some $m$ value occurs at equilibrium is
proportional to $W(N-m)$, where $W(n)$ is the statistical weight of
the atomic system defined in (\ref{Wofn}) (see Appendix B of
\inlinecite{arnaudAJP}).  As an example, consider two (distinguishable)
atoms ($N$=2).  A microstate of the isolated (matter$+$field) system
is specified by telling whether the first and second atoms are in
their upper (1) or lower (0) states and the value of $m$.  Since the
total energy is $N=2$, the complete collection of microstates (first
atom state, second atom state, field energy), is: (1,1,0), (1,0,1),
(0,1,1) and (0,0,2).  Since these four microstates are equally likely,
the probability that $m=0$ is proportional to 1, the probability that
$m=1$ is proportional to 2, and the probability that $m=2$ is
proportional to 1.  This is in agreement with the fact stated earlier
that $P(m)$ is proportional to $W(N-m)$.  After normalization, we
obtain for example that P(0)=1/4.
 
The normalized probability reads in general 
\begin{equation}\label{P}
    P(m)=\frac{W(N-m)}{Z}=\frac{N!}{2^{N}m!(N-m)!} .
\end{equation}
It is shown in the Appendix that the system entropy $S(t)$ increases
from $S(0)=\ln[W(N)]=0$ at the initial time ($m=0, n=N$), to
$S(\infty)=\ln[Z]=N \ln(2)$ when the equilibrium state has been
reached.  Note that no entropy is ascribed to single-mode fields, as
they are fully characterized by their energy.  The moments of $m$ are
defined as usual as
\begin{equation}\label{mr}
     \ave{m^{r}}\equiv \sum_{m=0}^{N}m^{r} P(m) ,
\end{equation}
where brackets denote averagings.  It is easily shown from (\ref{P}),
(\ref{mr}) that $\ave{m}=N/2$ and $\mathrm{var}(m) \equiv
{\ave{m^{2}}-\ave{m}^{2}} = N/4$.  Thus the number $m$ of
light quanta in the cavity fluctuates, but the statistics of $m$ is
sub-Poissonian, with a variance less than the mean.

The expression of $P(m)$ in (\ref{P}) just obtained has physical and
practical implications.  Suppose indeed that the equilibrium cavity
field is allowed to escape into free space, thereby generating an
optical pulse containing $m$ quanta.  It may happen, however, that no
pulse is emitted when one is expected, causing a counting error.  From
the expression in (\ref{P}) and the fact that $\ave{m}=N/2$, the
probability that no quanta be emitted is seen to be
$P(0)=4^{-\ave{m}}$.  For example, if the average number of light
quanta $\ave{m}$ is equal to $20$, the communication system
suffers from one counting error (no pulse received when one is
expected) on the average over approximately $10^{12}$ pulses.  Light
pulses of equal energy with Poissonian statistics are inferior to the
light presently considered in that one counting error is recorded on
the average over $\exp(\ave{m})=\exp(20)\approx 0.5~10^9$ pulses
(see, for example, p.~276 of \inlinecite{milonni}).

\section{Time evolution of the number of light quanta in isolated
cavities}
\label{sec:3}

Let us now evaluate the probability $P(m,t)$ that the number of light
quanta be $m$ at time $t$.  Note that here $m$ and $t$ represent two
independent variables.  A particular realization of the process was
denoted earlier $m(t)$.  It is hoped that this simplified notation
will not cause confusion.
 
Let $E(m)dt$ denote the probability that, given that the number of
light quanta is $m$ at time $t$, this number jumps to $m+1$ during the
infinitesimal time interval [$t, t+dt$], and let $A(m)dt$ denote the
probability that $m$ jumps to $m-1$ during that same time interval
(the letters ``$E$'' and ``$A$'' stand respectively for ``emission''
and ``absorption''). $P(m,t)$ obeys the relation
\begin{eqnarray}
    P(m,t+dt) & = & P(m+1,t)A(m+1)dt \nonumber\\
    & &+P(m-1,t)E(m-1)dt \nonumber \\
	& &+P(m,t) \left( 1 - A(m) dt - E(m) dt \right) .
	\label{FP}
\end{eqnarray}
Indeed, the probability of having $m$ quanta at time $t+dt$ is the sum
of the probabilities that this occurs via states $m+1$, $m-1$ or $m$
at time $t$.  All other possible states are two or more jumps away
from $m$ and thus contribute negligibly in the small $dt$ limit (see
\inlinecite{gillespie}, p.~381).  After a sufficiently long time, one expects
$P(m,t)$ to be independent of time, that is: $P(m,t+dt)=P(m,t)\equiv
{P(m)}$.  It is easy to see that (\ref{FP}) satisfies this condition
if
\begin{equation}\label{bal}
    P(m+1)A(m+1)=P(m)E(m) .
\end{equation}
This ``detailed balancing'' relation holds true because $m$ cannot go
negative (see \inlinecite{gillespie}, p.~425).  When the expression of
$P(m)$
obtained in (\ref{P}) is introduced in (\ref{bal}), one finds that
\begin{equation}\label{EA}
    \frac{E(m)}{A(m+1)}=\frac{P(m+1)}{P(m)}=\frac{N-m}{m+1} ,
\end{equation}
a relation that admits the  solution
\begin{equation}\label{Ei}
    E(m) = (N-m)(m+1) , \qquad
    A(m) = m^2 .
\end{equation}
    
It is natural to suppose that the probability $E$ of atomic decay is
proportional to the number $n$ of atoms in the upper state, and that
the probability $A$ of atomic promotion is proportional to the number
$N-n$ of atoms in the lower state.  Thus, we introduce the functions
of two variables ($n,m$)
\begin{equation}\label{Ei2}
    E(n,m) = n(m+1) , \qquad
    A(n,m) = (N-n) m ,
\end{equation}    
with the understanding that $E(m)=E(N-m,m)$ and $A(m)=A(N-m,m)$. 
These relations hold within a proportionality factor.  Setting this
proportionality factor as unity amounts to fixing the time scale.  The
expressions in (\ref{Ei2}) say that the probability that an atom gets
promoted to the upper level in the time interval $[t,t+dt]$ is equal
to $mdt$, while the probability of atomic decay is $(m+1)dt$.  These
expressions were obtained by Einstein in 1917 in a somewhat different
manner~\endnote{Assuming that the atoms emit or absorb a single light
quantum at a time (``1-photon'' process), the strictly-classical limit
tells us that the probability that an atom in the upper state decays
must be a linear function of $m$, i.e., we must have $E(n,m)=n(am+b)$,
where $a,b$ and later $c,d$ are constants.  Likewise, the probability
of atomic promotion must be of the form: $A(m)=(N-n)(cm+d)$.  But,
furthermore, $A$ is required to vanish for $m=0$ since, otherwise, $m$
could go negative.  Accordingly, $d=0$.  Remembering that $n=N-m$, we
find upon substitution of $E(m)=(N-m)(am+b)$ and $A(m)=mcm$ in the
detailed-balancing relation and simplifying that: $am+b=cm+c$, a
relation which is satisfied for all $m$ values only if $a=c$ (equality
of stimulated emission and absorption coefficients), and $b=c=a$.  To
within a constant factor, we have therefore $E(n,m)=n(m+1)$, and
$A(n,m)=(N-n)m$, relations discovered by Einstein at the turn of the
previous century.  Note that the ``1'' in the term $m+1$ of $E(n,m)$
is sometimes ascribed to spontaneous emission in the mode.  The lack
of symmetry between $m+1$ (emission) and $m$ (absorption) is only
apparent.  If indeed the field energy is defined as $m+1/2$, upward
and downward transition probabilities may be both written as the
arithmetic averages of the field energy before and after the jump.}.

Let us now restrict our attention to the steady-state regime and large
values of $N$.  Since $m$ is large, the ``1'' in the expression $m+1$
of $E(m)$ may be neglected.  Furthermore, in that limit, $m$ may be
viewed as a continuous function of time with a well-defined
time-derivative.  Because the standard deviation $\sqrt{N/4}$ of $m$
is much smaller than the average value, the so-called ``weak-noise
approximation'' is permissible \cite{gillespie}.  Within that
approximation, the average value $\ave{f(n,m)}$ of any smooth function
$f(n,m)$ may be taken as approximately equal to $f(\ave{n},\ave{m})$.
  
The evolution in time of a particular realization $m(t)$ of the
process obeys the classical Langevin equation
\begin{equation}\label{L1}
    \frac{dm}{dt}= \mathcal{E}-\mathcal {A} ,
\end{equation}
where
\begin{equation}\label{aux}
   {\mathcal{E}} \equiv E(m) + e(t) , \qquad
   {\mathcal{A}} \equiv A(m) + a(t) .
\end{equation}    
In these expressions, $e(t)$ and $a(t)$ represent uncorrelated
white-noise processes whose spectral densities equal to $E\equiv
E(\ave{m})$ and $A\equiv A(\ave{m})$, respectively~\endnote{A formal
proof of the validity of the Langevin equation will not be given here. 
Instead, it will be shown that the variance of $m$ obtained from the
Langevin equation coincide with the result obtained directly from
Statistical Mechanics}.  Without the noise sources, the evolution of
$m$ in (\ref{L1}) would be deterministic, with a time-derivative
equals to the drift term $E(m)-A(m)$.  If the expressions (\ref{Ei})
are employed, the Langevin equation (\ref{L1}) reads
\begin{eqnarray}
    \frac{dm}{dt} & = & N m - 2 m^2 + e - a , \nonumber\\
	S_{e-a} & = & E+A = N \ave{m} = N^{2}/2 ,
	\label{L3}
\end{eqnarray}
where the approximation $N \gg 1$ has been made.
  
Let $m(t)$ be expressed as the sum of its average value $\ave{m}$ plus
a small deviation $\Delta m(t)$, and $Nm-2m^{2}$ in (\ref{L3}) be
expanded to first order.  A Fourier transformation of $\Delta m(t)$
with respect to time amounts to replacing $d/dt$ by
$j\Omega$~\endnotemark[4].
The Langevin equation now reads
\begin{equation}\label{L4}
    j \Omega \Delta m = -N \Delta m+e-a , \qquad
    S_{e-a} = N^{2}/2 ,
\end{equation}
where $m$ has been replaced by its average value $N/2$. 

Since the spectral density of $z(t)=ax(t)$, where $a\equiv{a'+ja''}$
is a complex number and $x(t)$ a stationary process, reads:
$S_{z}(\Omega)=|a|^2 S_{x}(\Omega)$, one finds from (\ref{L4}) that
the spectral density of the $\Delta m(t)$ process is
\begin{equation}\label{sp}
    S_{\Delta m}(\Omega)=\frac{N^{2}/2}{N^{2}+\Omega ^{2}} .
\end{equation} 
The variance of $m$ is the integral of $S_{\Delta m}(\Omega)$ over
frequency ($\Omega/2\pi$) from minus to plus infinity, that is:
var$(m)=N/4$ in agreement with the previous result in Section 2.
 
Suppose now that a small absorbing body, perhaps a single atom that
remains most of the time in its lower state as discussed earlier, is
introduced in the cavity.  One expects that this unique atom will not
affect significantly the average value and the statistics of $m$ for
some period of time.  If $m$ were non-fluctuating, the probability that
a detection event occurs during the time interval $[t,t+dt]$, divided
by $dt$, would be a constant.  This property defines the Poisson
process.  Since $m$ actually suffers from the fluctuations discussed
earlier in this section, the detection rate is super-Poissonian, with
a spectral density that exceeds the average detection rate.
 
The expression for the detection rate has the same form as the one
introduced earlier for the stimulated absorption rate $A$, the only
difference being that in the detector atoms remain in the lower state
most of the time, as discussed in the introduction. Accordingly 
\begin{equation}\label{Q}
    {\mathcal{Q}} = \alpha m + q , \qquad
    \Delta Q = \alpha \Delta m + q ,
\end{equation} 
where $\alpha$ denotes a constant.  The noise sources $e(t)$, $a(t)$,
and $q(t)$ are uncorrelated white-noise processes of spectral
densities $E$, $A$, and $Q\equiv \alpha \ave{m}$, respectively.

The spectral density of the detection rate fluctuation $\Delta Q$
defined in (\ref{Q}) may be obtained directly from (\ref{sp}) since
$q(t)$ is presently supposed to be uncorrelated with the other noise
sources
\begin{equation}\label{det}
   S_{\Delta Q}(\Omega)=\alpha^2 S_{\Delta 
   m}(\Omega)+Q=\alpha^{2}\frac{N^2/2}{N^2+\Omega^2}+Q .
\end{equation} 

Consideration of optical cavities close to a state of equilibrium
therefore provides some information concerning the detection rate
statistics.  But the conclusion that the detection rate fluctuations
always exceed the shot-noise level holds true only when $\Omega \gg
\sqrt {\alpha N} $.  After a sufficiently long time, even a single
absorbing atom affects the statistics of $m$ in such a drastic way
that the result in (\ref{det}) becomes invalid.  A true steady-state
may be obtained only if a pumping mechanism compensates for the energy
loss caused by light-quanta absorption.  The accurate result, to be
given next, shows that, at low frequencies, the detection rate depends
in a crucial way on the pump fluctuations.

\section{Laser noise}
\label{sec:4}

Lasers are open systems with a source of energy called the pump, and a
sink of energy presently viewed as an ideal optical detector.  It is
natural to suppose that the probabilities of atomic decay or atomic
promotion that were found earlier consistent with the laws of
statistical mechanics, still hold when there is a supply of atoms in
the upper state (the pump), and an absorber of light energy (the
detector).

The evolution equation for the number $m$ of light quanta is thus
obtained by subtracting the loss rate $\mathcal{Q}$ from the
right-hand-side of (\ref{L1}).  Since the system is not isolated,
$n+m$ may now fluctuate, and the expressions of $E(n,m)$ and $A(n,m)$
given in (\ref{Ei2}) must be employed.  A second equation describing
the evolution of the number $n$ of atoms in the upper state is needed,
which involves the prescribed pump rate ${\mathcal{J}}(t)$.
To summarize, the evolution equations for $m$ and $n$ are
\begin{eqnarray}
    \frac{dm}{dt} & = & \mathcal{E}-\mathcal{A}- \mathcal{Q} ,
    \label{Laserm}  \\
    \frac{dn}{dt} & = & \mathcal{J}-\mathcal{E}+\mathcal{A} ,
    \label{Lasern}
\end{eqnarray}
where
\begin{equation}
	\begin{array}{lll}
		{\mathcal{E}} \equiv E(n,m)+e(t) , \quad & E(n,m) = n m , \quad & S_{e} = E , \\
		{\mathcal{A}} \equiv A(n,m)+a(t) , \quad & A(n,m) = (N-n) m , \quad & S_{a} = A , \\
		{\mathcal{Q}} \equiv \alpha m+q(t) , \quad &  & S_{q} = Q , \\
		{\mathcal{J}} \equiv J+\Delta J(t) . &  & 
	\end{array}
\end{equation}

In the steady-state, the right-hand-sides of (\ref{Laserm}) and
(\ref{Lasern}) vanish, and we have: $J=E-A=Q$, that is
$J=(2\ave{n}-N)\ave{m}=\alpha \ave{m}$.  Thus,
$\ave{m}=J/\alpha$ and $2\ave{n}-N=\alpha$.  This relation
expresses the fact that the stimulated emission gain coefficient
($\ave{n}$) minus the stimulated absorption loss coefficient
($N-\ave{n}$) equals in the steady state the linear loss
coefficient $\alpha$.

Next, observe that at small frequencies, the left-hand-sides of the
previous equations (\ref{Laserm}) and (\ref{Lasern}) may be neglected. 
The simple relation ${\mathcal{Q}}(t)={\mathcal{J}}(t)$ follows,
proving that the detection rate does not fluctuate
(${\mathcal{Q}}=$constant) if the pump is non-fluctuating or
``quiet'' (${\mathcal{J}}=$constant).  The relation
${\mathcal{Q}}(t)={\mathcal{J}}(t)$ holds at low frequencies even in
the presence of internal \emph{conservative} effects such as gain
compression (due, for example, to spectral hole burning), gain
guidance, or mesoscopic effects that occur when the thermal energy
$k_{B}T$ is not large compared to the average level spacings
\cite{arnaudAJP,arnaudPRB}.

A quiet pump is henceforth assumed.  When the above equations are
linearized and $\Delta{m}, \Delta{n}$ are Fourier transformed, one
obtains
\begin{eqnarray}
    j \Omega \Delta m & = & 2 \ave{m} \Delta n + e - a - q ,
    \label{Lm}  \\
    j \Omega \Delta n & = & -2 \ave{m} \Delta n - \alpha \Delta m - e + 
    a .
    \label{Ln}
\end{eqnarray}
Let us recall that $e$, $a$ and $q$ are uncorrelated white-noise
processes whose spectral densities are equal to the corresponding
average rates.  After elimination of $\Delta n$ from the above two
equations, $\Delta m$ may be expressed in terms
of uncorrelated noise sources 
\begin{equation}\label{deltam}
	\Delta m=\frac{j\Omega(e-a)-(2\ave{m}+j\Omega)q} {j\Omega
	2\ave{m}+\alpha 2\ave{m}-\Omega^2} .
\end{equation}

We then proceed as in the previous section, evaluating the spectral
density of $\Delta m$ and integrating over frequency to obtain the
variance of $m$. The result is 
\begin{equation} \label{varm}
	\frac{\mathrm{var}(m)}{\ave{m}} =
	\frac{N+\alpha}{4\ave{m}}+\frac{1}{2} .
\end{equation}
In the limit that $\alpha \ll N$ and $\ave{m}=N/2$, the
right-hand-side of (\ref{varm}) is 1 while the corresponding result in
Section \ref{sec:2} relating to the isolated cavity is 1/2.  This is
due to the singular behavior of the spectral density of $\Delta m$ at
$\Omega=0$ in the limit considered.  Physically, this means that small
losses allow $m(t)$ to drift slowly.

We are mostly interested in the detection rate fluctuation $\Delta
Q=\alpha \Delta m + q$.  Notice that $m$ and $q$ \emph{are}
correlated.  Proceeding as in the previous section, we obtain
\begin{equation}\label{X}
	\frac {S_{\Delta Q}(\Omega)}{Q}-1=\frac{\left(
	(N+\alpha)/4\alpha \ave{m}^{2} \right) \Omega
	^2-1}{(\Omega/\alpha)^2+ \left( 1-\Omega^2/2\alpha \ave{m}
	\right) ^2} .
\end{equation}
In the limit that $\alpha \ll N$, $\ave{m}=N/2$, the above result
reduces to the one given in Section \ref{sec:3} at high frequencies:
$\Omega \gg \sqrt {\alpha N}$.

\begin{figure}
    \centering
    \includegraphics[scale=0.7]{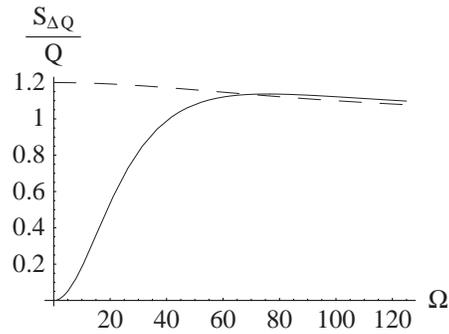}
	\caption{Normalized spectral density $S_{\Delta Q}(\Omega )/Q$
	of the detection rate $\mathcal{Q}$ as a function of the
	normalized baseband (or Fourier) frequency $\Omega$.  The
	number of atoms is $N=100$ and the cavity loss rate (due to
	the detection process) is $\alpha=20$.  The average number of
	light quanta is $\ave{m}=N/2$.  Plain lines: exact result from
	the discussion in Section \ref{sec:4}.  Dashed lines:
	approximate result for nearly isolated cavities, see Section
	\ref{sec:3}.}
    \label{fig:2}
\end{figure}

Comparison between the results in (\ref{det}) and (\ref{X}) is
exemplified in Figure \ref{fig:2} for $N=100$, $\alpha=20$ and
$m=N/2$.  In the exact treatment, the spectral densities of the
photodetection process go to zero at zero baseband (or Fourier)
frequencies.  Note also that, for the parameter values considered, a
relaxation oscillation peak appears.  In the large optical power limit
($\ave{m} \gg 1$), the above expression reduces to
\begin{equation}\label{Xbis}
	\frac {S_{\Delta
	Q}(\Omega)}{Q}=1-\frac{1}{(\Omega/\alpha)^2+1} .
\end{equation}

Expressions obtained from the present theory have invariably been
found to coincide with the Quantum Optics results when the same
approximations are made, essentially the large atom number
approximation.  In particular, the simple expression in
(\ref{Xbis}) was first given by \inlinecite{yamamoto}, see Figure
15-10, b).  The expression for the atomic-number detection-rate
correlation was first obtained from a theory similar to the one
described in the present paper \cite{arnaudIEEJ}.

\section{Conclusion}

The purpose of this paper was to show that, in the limit of a large
number of atoms, important results relating to laser light
fluctuations, usually derived on the basis of Quantum Optics, may be
obtained in a much simpler manner.  This is so even if the detected
light statistics is sub-Poissonian.  The light field enters only
through its energy, which is quantized as a result of atomic
quantization and conservation of energy but not directly.

For the sake of simplicity, many effects have been neglected.  But the
theory may be generalized to account for spontaneous carrier
recombination, phase-amplitude coupling, and complicated cavity
structures \cite{siegman,siegmanAPB}.  Besides the intensity noise,
other useful quantities may be evaluated, particularly phase noise
\cite{arnaudOL,arnaudQSE}.  When the atoms get close to one another,
the upper and lower levels spread into bands called in the field of
Semiconductor Physics, ``conduction'' and ``valence'' bands, with
$k_{B}T$ small compared with the widths of the bands.  In that
situation the Fermi-Dirac distribution (see for example
\inlinecite{chusseau}) is applicable.  This involves large changes in
the parameter values in comparison with those given in the present
paper, but the general principles remain the same.

\appendix
 
\section{Field entropy}

The purpose of this appendix is to show that the present theory
predicts an increase of the system (single-mode cavity+atoms) entropy
in the course of time, as is required by the second law of
thermodynamics.  Since the single-mode field is entirely defined by
only one parameter, namely its energy, its entropy vanishes.  When
atoms are introduced in an empty cavity, the matter energy decreases,
part of it being converted into field energy.  The system entropy
nevertheless \emph{increases} in the course of time because the number
of energy states available to matter increases.
 
Let $W(n)$ denote the statistical weight of a collection of $N$ atoms,
with energy $n$ (number of atoms in the upper state), as given in
(\ref{Wofn}) of the main text.  If the $N$ atoms are introduced in their
upper state into an empty cavity, initially ($t=0$), the matter
statistical weight $W(N)=1$ and the matter entropy $\ln[W(N)]$
vanishes.  The system entropy $S(0)$ vanishes also since no entropy is
ascribed to the field.
   
Suppose now that the system has reached a state of equilibrium
(formally, $t=\infty$).  When an isolated system of total energy $N$
consists of two parts, one with energy $n$ and statistical weight
$W(n)$, and the second with energy $m$ and statistical weight $W'(m)$,
with $n+m=N$, its statistical weight
reads (see p.~15 in \inlinecite{kubo})
\begin{equation}\label{entrop}
    W_{system}(\infty)=\sum_{n=0}^{N}W(n)W'(N-n) .
\end{equation}

In the present situation, the (single-mode) field statistical weight
$W'(m)$ is unity, and therefore the system entropy simplifies to
\begin{eqnarray}
    S(\infty) & \equiv & \ln({W_{system}(\infty)}) \nonumber\\
    & = & \ln \left( \sum_{n=0}^{N}W(n) \right) \nonumber\\
    & = & \ln(Z) \nonumber\\
    & = & N\ln(2) ,
    \label{s-e}
\end{eqnarray}
if the expression of $Z$ given in (\ref{Z}) is used.  Thus the system
entropy increases with time as asserted earlier.

For a total system energy $U\le N$, the probability $P(m)$ of having
$m$ light quanta in the cavity is, more generally
\begin{equation}
    P(m) \propto \frac {N!}{(U-m)!(N-U+m)!} ,
    \label{lin}
\end{equation}
where (\ref{Wofn}) and the relation $n+m=U$ have been used.  When $U$
is somewhat less than $N/2$ (precisely $N-2U\gg1$), first-order
expansion of $\ln \left( P(m) \right)$ shows that $P(m)$, as given in
(\ref {lin}) is almost proportional to $\exp(-\beta m)$, where the
Boltzmann factor $\exp(-\beta )=\ave{n}/(N-\ave{n})$.  This is
essentially the thermal regime considered by Planck and Einstein at
the turn of the previous century.

To make contact with Quantum Optics concepts, let us show that the
entropy that Quantum Optics assign to single-mode fields is the
difference between the system entropy and the average matter entropy. 
We have indeed the mathematical identity
\begin{eqnarray}
    S & = & \ln(Z) \nonumber\\
    & = & \sum {\textstyle \frac{W(n)}{Z}} \ln(Z) \nonumber\\
    & = & \sum {\textstyle \frac{W(n)}{Z}} \ln(W(n)) - \sum
    {\textstyle \frac{W(n)}{Z}} \ln(W(n)/Z) ,
    \label{id}
\end{eqnarray}
where the sums are from $n=0$ to $N$.  The entropy of matter is $\ln
\left( W(n) \right)$ if its energy is known to be $n$.  The
probability that some $n$ value occurs in the cavity is $W(n)/Z$. 
Accordingly, the first term in the final expression of (\ref {id}) is
recognized as the average matter entropy.
 
On the other hand, the second term on the right-hand-side of
(\ref{id}) may be written as
\begin{equation}\label{e-field}
    S_{field}=-\sum_{m=0}^{N} P(m)\ln \left( P(m) \right) ,
\end{equation}
where the summation over $n$ has been replaced by a summation over
$m$, and we have defined, as in the main text, $P(m)=W(N-m)/Z$. 
Equation (\ref {e-field}) is the standard expression of field entropy. 
For $U=N=1000$ atoms, for example, one calculates from the above
expressions that the system entropy $S=693.15$ may be split into an
average matter entropy $S_{matter}=688.97$ and a field entropy
$S_{field}=4.18$.  On the other hand, in the limit $U\ll{N}$, the
expression in (\ref {e-field}) coincides with the expression obtained
from the Quantum Optics method that treats single-mode fields as
quantized harmonic oscillators in contact with a thermal bath at
temperature reciprocal $\beta$.

\theendnotes

\begin{acknowledgements}
    The author wishes to express his thanks to L. Chusseau, F.
    Philippe and anonymous referees for helpful observations.
\end{acknowledgements}

\end {document}